\documentclass[useAMS,usenatbib]{mn2e} 

\usepackage{times,aas_macros,multirow,natbib}
\usepackage{epsfig}
\usepackage{graphicx}
\usepackage{amssymb}

\begin{document}
\title[A catalogue of Stripe 82 clusters]{4098 galaxy clusters to
  z$\sim$0.6 in the Sloan Digital Sky Survey equatorial Stripe 82}
\author[Geach, Murphy \& Bower]
{\parbox[h]{\textwidth}{James~E.~Geach,$^{1,2}$\thanks{E-mail:
      jimgeach@physics.mcgill.ca} David\ N.\ A.\ Murphy$^{2}$ \& Richard\ G.\
    Bower$^{2}$}
  \vspace*{3pt}\\
  \noindent $^1$Department of Physics, McGill University, Ernest
  Rutherford Building, 3600 Rue University, Montr\'eal, Qu\'ebec, H3A
  2T8, Canada. \\
  \noindent $^2$Institute for Computational Cosmology, Durham University, South Road, Durham. DH1 3LE, UK.\\}

\pagerange{\pageref{firstpage}--\pageref{lastpage}}\pubyear{2011}

\maketitle

\label{firstpage}

\begin{abstract} We present a catalogue of 4098 photometrically selected
galaxy clusters with a median redshift $\left<z\right>=0.32$ in the 270\
square degree `Stripe 82' region of the Sloan Digital Sky Survey (SDSS),
covering the celestial equator in the Southern Galactic Cap
($-50^\circ<\alpha<59^\circ$, $|\delta|\leq1.25^\circ$). Owing to the
multi-epoch SDSS coverage of this region, the {\it ugriz} photometry is
$\sim$2\,magnitudes deeper than single scans within the main SDSS footprint.
We exploit this to detect clusters of galaxies using an algorithm that
searches for statistically significant overdensities of galaxies in a Voronoi
tessellation of the projected sky. 32\% of the clusters have at least one
member with a spectroscopic redshift from existing public data (SDSS Data
Release 7, 2SLAQ \& WiggleZ), and the remainder have a robust photometric
redshift (accurate to $\sim$5--9\% at the median redshift of the sample). The
weighted average of the member galaxies' redshifts provides a reasonably
accurate estimate of the cluster redshift. The cluster catalogue is publicly
available for exploitation by the community to pursue a range of science
objectives. In addition to the cluster catalogue, we provide a linked
catalogue of 18,295 $V\leq21$\,mag quasar sight-lines with impact parameters
within $\leq$3\,Mpc of the cluster cores selected from the catalogue of Veron
et al.\ (2010). The background quasars cover $0.25 < z < 2$, where Mg\,{\sc
ii} absorption-line systems associated with the clusters are detectable in
optical spectra. \end{abstract} 

\section{Introduction}

Efficient, reliable galaxy cluster detection is a long-standing, and
perhaps clich\'ed, astronomical problem. However, as we move into the
era of large scale, `petabyte' sky surveys, the issue is especially
pertinent. The practical uses of groups and clusters are very well
known. Since they betray the presence of underlying dark matter
potentials, clusters' abundance and distribution are probes of
primordial fluctuations in dark matter density and its subsequent
growth. The sensitivity of clusters for use as probes of large scale
structure is high because they probe the exponential tail
of the mass distribution; thus clusters can provide cosmological
constraints on the nature of dark energy and test the assumption that
the primordial density field has a Gaussian distribution (Gladders et
al.\ 2007; Rozo et al.\ 2010). Cross-correlating the positions of
clusters with fluctuations in the microwave background provides a
direct test of the expansion rate of the Universe through the
integrated Sachs-Wolfe effect (Ho et al.\ 2008; Sawangwit et al.\
2010).  The galaxies that occupy clusters represent an important
sample for studies of galaxy evolution. It has long been known that
galaxies' environments have a profound influence on their star
formation histories: the galaxies in the cores of rich clusters tend
to be passive in terms of their current star formation activity.
Identifying and correctly modelling the mechanisms that drive this
behaviour presents an important challenge for galaxy formation models
(McCarthy et al. 2008, Kapferer et al.\ 2008), and casts important
light on the recent decline in the cosmic star formation rate.

Fortuitously, it is this characteristic of galaxies in rich clusters
that aids in their detection against myriad background and
foreground galaxies projected onto the celestial sphere. As galaxies
accumulate in the potential wells of clusters, their star formation
rates are curtailed (whether this happens gradually or rapidly is a
matter of some debate). The passively evolving stellar populations
develop strong metal absorption lines blue-ward of 4000\AA, giving
rise to a break (colloquially, `the 4000\AA\ break') in their
spectra. Cluster members therefore appear red in broad-band filters
that straddle this feature. Since galaxies in clusters cover a
range in mass, the combination of this characteristic color and range
of luminosities form a distinct ridge or sequence in colour-magnitude
parameter space. Star-forming galaxies in the outskirts of the cluster
(the `blue cloud') are thought to eventually have their star formation
truncated by environmental processes or terminated by gas exhaustion;
subsequent passive evolution enables these galaxies to `pile up' on
the red sequence. One can select for galaxies in this narrow colour
range (there can also be some luminosity-dependent tilt in the
sequence, caused by metallicity or age effects), to isolate galaxies
belonging to the cluster (Gladders \& Yee\ 2000, 2005). Due to the
redshift, the red-sequence is
detected in ever redder filter combinations, and so in a deep
panchromatic survey one can use a simple combination of filters to
isolate clusters as a function of epoch.

Many cluster finding methods exist, but this article describes a
generic algorithm for detecting over-densities in a panoramic
photometric survey. It has been designed specifically for use with the
Panoramic Survey Telescope and Rapid Response System
(PanSTARRS\footnote{\tt http://pan-starrs.ifa.hawaii.edu}) survey, but
is applicable to any set of photometric data. The first of four 1.8\,m
PanSTARRS telescopes is located on Haleakala, Hawaii. Its 1.4
gigapixel camera images $\sim$7 square degrees per shot, and will
continuously scan in {\it grizy} filters, encompassing the entire
night sky visible from Hawaii once every (dark) lunar cycle.  Over
three years of operations, PS1 will build up its $r<24$\,mag 3$\Pi$
survey, including a deeper $r<27$\,mag `Medium Deep Survey' (MDS) over
$\sim$80 square degrees. In lieu of PanSTARRS data (which at the time
of writing is being accumulated; survey mode commenced in May 2010),
we have put our algorithm to immediate use on another existing public
imaging survey -- the Sloan Digital Sky Survey (SDSS; York et al.\
2000; see Abazajian et al.\ 2009 for details on the 7th data release
[DR7]). In particular, we have concentrated our efforts on a specific
sub-region within the SDSS which was re-visited many times -- a deeper
equatorial strip, $|\delta|\leq1.25^\circ$, spanning $-50^\circ<\alpha <59^\circ$
in right ascension, and totalling approximately 270 square
degrees. The co-addition of 47 and 55 strip scans (corresponding to
the southern and northern parts of the Stripe) results in a catalogue
of objects which probes $\sim$2 magnitudes deeper than an individual scan.

In this paper we present a catalogue of $\sim$4100 clusters detected
in the stripe, including a brief description of the algorithm
itself. An exhaustive description of the full workings of the cluster
detection method, along with statistical tests on sophisticated `mock'
catalogues will be presented in Murphy, Geach \& Bower\ (2010). In \S2
we describe the algorithm, in \S3 we describe the application to 
SDSS Stripe 82 and the resulting cluster catalogue, including vital statistics such as
redshift distribution and richness estimates. We also present a
supplementary catalogue of $\sim$18,300 QSO sight-lines which pass
within 3 proper Mpc of the cluster centres. The sight-line catalogue
is constructed from the catalogue of Veron et al. (2010) and could be
a useful resource for future absorption line studies which aim to
probe the properties of the intracluster medium, and the gas
properties of galaxies in rich environments. Finally, we include an
Appendix describing how readers can access and use the
cluster/sight-line catalogues. Throughout -- unless otherwise stated in
the text -- we will be referring to photometry on the Sloan system
(Gunn et al.\ 1998), and -- where relevant -- assume a cosmology with
$H_0=70$\,km\,s$^{-1}$\,Mpc$^{-1}$, $\Omega_{\rm m}=0.3$ and
$\Omega_{\rm \Lambda}=0.7$.

\section{The method}

Our cluster detection algorithm is generic, in that it can be applied
to any wide-area photometric data, including extension into the near-
and mid-infrared. The main aim is to efficiently identify
overdensities in projection, with multi-band photometry helping to
isolate structures at specific redshifts. An exhaustive description of
the cluster detection algorithm, including tests on mock catalogues is
described in Murphy, Geach \& Bower\ (2010), however in this section
we outline the principle elements of the cluster finder.

\subsection{Identifying overdensities with Voronoi tessellation}

In large scale imaging surveys, physical associations of galaxies will
manifest themselves as overdensities in the projected `field' of
foreground and background galaxies. The contrast (and thus
detectability) of such structures can be enhanced by applying simple
selections in luminosity and colour, since cluster of galaxies tend to
be dominated by a population of passive galaxies with a narrow
distribution of red colours. The evolution of the expected colours of
this `red sequence' is reasonably well modelled from simple stellar
populations, and so one can attempt to isolate clusters of galaxies as
a function of redshift.
 
Having reduced the contamination of background and foreground galaxies
with colour selections, one must identify associations of galaxies --
i.e. group them into clusters. This requires estimating the local
surface density, $\Sigma$, and linking galaxies that reside in a
common region of enhanced density. The optimal way to assess whether
an individual galaxy resides in an overdensity is compare the
probability of finding a galaxy with $\Sigma$ in a random field.  To
estimate $\Sigma$ for a given galaxy, we calculate Voronoi diagram for
the 2D galaxy distribution The Voronoi diagram is a tessellation of
convex hulls -- `cells' -- with each galaxy occupying exactly one
cell. The set of co-ordinates within a galaxy's cell are closer to
that galaxy than any other. The inverse of the area of the cell is the
optimal estimate of the local $\Sigma$. Percolating adjacent cells
satisfying some threshold criterion allows one to systematically
detect statistically significant structures. Voronoi tessellation has
been used in other cluster/structure finding algorithms (e.g. van
Breukelen \& Clewley\ 2009; Ramella et al.\ 2001; Ebeling \&
Wiedenmann\ 1993). An advantage of this method compared to some other
cluster detection techniques is that it makes no assumptions about the
shape of the overdensity, allowing one to search for extended
filamentary structures, as well as regular virialised systems. Indeed,
one need not group galaxies into discrete clusters, but simply use the
Voronoi tessellation to produce maps of local surface density.

The statistical significance of finding a cell with area $a$ can be
obtained by comparing to the probability of finding a cell with this
area in a randomly distributed catalogue. This has been shown to
approximate to the Kiang distribution (Kiang\ 1966):
\begin{equation}
P(a) = \int_0^a {\rm d}p = 1 -  e^{-4a}\left(\frac{32a^3}{3} + 8a^2 + 4a + 1 \right)
\end{equation}
Where the area $a$ (calculated by triangulating the convex hull) is
normalised to the average area cell area (in the random
catalogue). Galaxies residing in overdense regions can be flagged
where $p_i < p_c$, with $p_c$ representing some critical probability
threshold. 

\subsection{The detection algorithm}

A more comprehensive description of the algorithm is given in Murphy,
Geach \& Bower (2010), but it is instructive to give a brief overview
here.  In summary, the basic detection algorithm can be described as
follows:
\begin{enumerate}
\item[1] Apply a photometric cut to input catalogue (e.g. a simple
  colour cut, or more sophisticate photometric redshift selection).
\item[2] Calculate Voronoi diagram of real catalogue. For each galaxy
  calculate the probability that its Voronoi cell would be found in
  the random field, $p(a')$ (equation\ 1), where $a'$ is the
  normalised area: $a' = a/\bar{a}$. Here $\bar{a}$ is equivalent to
  the average galaxy surface density (in the photometric cut).
\item[3] Only considering galaxies with $p_i < p_c$, we move through
  the galaxy catalogue sorted in ascending $p$. Cells are percolated
  such that connected cells (i.e.\ those with shared Voronoi vertices)
  are assembled into putative clusters. Each time a galaxy is added,
  the average density of the `cluster' is assessed, and the
  percolation is terminated when the average density falls below 10
  times the average density of the field. The percolation also stops
  if no more cells with $p_i < p_c$ can be added to the
  conglomeration.
\item[4] Groups of connected cells are classified as clusters if they
  have $N\geq N_{\rm min}$ galaxies. We we choose $N_{\rm min}=5$ as a
  suitable value.
\end{enumerate}

Although this algorithm is generic in application to an arbitrary
selection method, in this work we will use simple linear colour
selections to help isolate clusters at specific redshifts. This relies
on the reddening of the linear `red-sequence' ridge prominent in
groups and clusters of galaxies, such that -- in principle --
repeatedly applying this algorithm over a wide range of colour
selections, one effectively `scans' over a range of redshifts. Note
that this technique makes few assumptions
about the nature of the clusters, aside from the fact that they are
densely packed associations of galaxies on the sky, and that a
significant fraction of galaxies in these associations have similar
colours.

\begin{figure*}
\centerline{\includegraphics[width=0.99\textwidth]{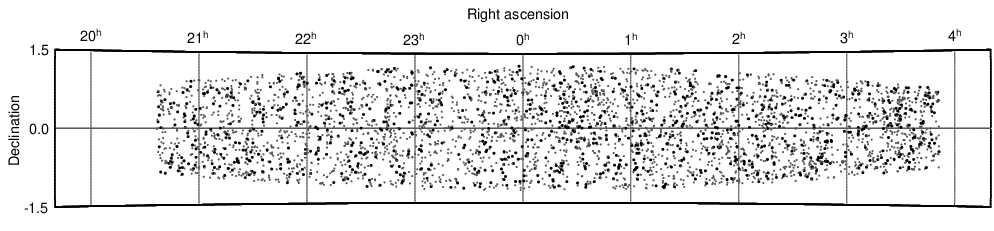}}
\centerline{\includegraphics[width=0.90\textwidth]{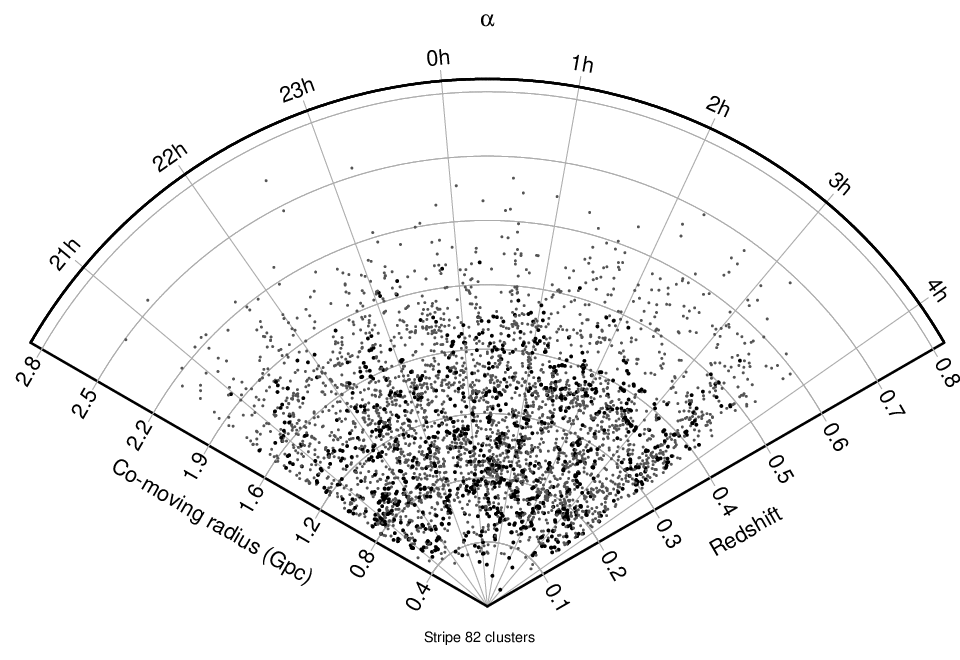}}
\caption{Sky and redshift distribution of clusters detected in Stripe
  82. The top panel shows the angular distribution (Aitoff projection)
  and the lower cone plot indicates the redshift distribution over the
  full right ascension range of the Stripe, projected in $\sim$2.5
  degrees of declination. Larger points indicate BCGs with
  spectroscopic measurements (we describe redshift estimation of
  clusters in \S\S3.3.1). We comfortably detect clusters out to
  $z\sim0.6$, with a handful of systems potentially detected at higher
  redshifts. Beyond this, the 4000\AA\ break moves into the near-IR,
  and selection using the Sloan optical bands alone becomes
  inefficient.}
\end{figure*}

\section{Detection of clusters in the Sloan Digital Sky Survey
  Equatorial Stripe 82}

\subsection{Data selection}

We used the SDSS Catalog Archive Server (CAS)\footnote{\tt http://casjobs.sdss.org} to
extract {\it griz} `{\tt modelMag}' photometry from the {\tt PhotoObj} table for all
galaxies in the Stripe 82 co-add. We enforce a magnitude limiting range: $14<
r\leq24$\,mag, and to eliminate stellar contamination we stipulate an additional
minimum offset between the point-spread function fit magnitude and model magnitude:
$(r_{\rm PSF} - r_{\rm model}) > 0.05$\,mag. All photometry is corrected for Galactic
extinction using the relevant `{\tt extinction}' table value (Schlegel et al.\ 1998).
To remove overly de-blended, saturated and sources near frame edges, we also make use
of the CAS {\tt fPhotoFlags} parameter. We require all of the following to hold:
\begin{itemize} \item {\tt BINNED1} or {\tt BINNED2} or {\tt BINNED4} $>$ 0 \item {\tt
BLENDED} or {\tt NODEBLEND} or {\tt CHILD} !$=$ {\tt BLENDED} \item {\tt EDGE} or {\tt
SATURATED} $=$ 0 \end{itemize} There are a total of 11,154,087 galaxies in the
catalogue, and for convenience we split them into 0.2$^{\rm h}$ sectors in right
ascension, overlapping by $\sim$3$^{\rm m}$. For every galaxy, we determine whether a
spectroscopic redshift is available from either the SDSS Data Release 7 (DR7),
2dF--SDSS LRG and QSO (2SLAQ; Croom et al.\ 2009) or WiggleZ Data Release 1
(Drinkwater et al.\ 2010). If no spectroscopic redshift is available, we ingest DR7
photometric redshifts (Abazajian et al.\ 2009). We discuss the photometric redshifts
in further detail in \S3.3.1.

\subsection{Cluster detection}

To elaborate on the colour scanning technique described in \S2.2, we apply the
Voronoi tessellation after selecting galaxies in narrow `slices' of colour in
the $(g-r)$, $(r-i)$ and $(i-z)$ bands. Each slice is defined by a linear
strip in colour-magnitude space, which can be normalised in colour (the
normalisation is defined as the colour at 20th magnitude), and has a gradient
and width. Although this filtering could be adapted or refined in several ways
(for example, allowing the gradient of the slice to vary), in this catalogue
we have chosen simply to apply a filter that fixes the width and slope for
slices in $(g-r)$, $(r-i)$ and $(i-z)$. We assume that the slope of the
red-sequence (i.e.\ where we expect the contrast of a cluster against the
background will be maximised in a colour-slice) is also constant with
redshift. This method of exploiting the red-sequence to detect clusters of
galaxies was pioneered by Gladders \& Yee (2000).

To fix the slope and width of the slices, we turn to the richest known cluster
in Stripe 82: Abell\ 2631 (Abell et al.\ 1989; B\"ohringer et al.\ 2000). We
linearly fit the colour-magnitude sequence in each set of filters for 126
cluster members. The slopes of the colour-magnitude relation in $(g-r)$,
$(r-i)$ and $(i-z)$ are $-0.048$, $-0.017$ and $-0.023$ respectively. The
width of each slice is increased until it selects 90\% of the members (see
Gladders et al.\ 1998), and in the same bands we find that the required widths
are 0.152, 0.067 and 0.110\,mag. We fixed the width of all slices to the
largest of these, although some refinement or adaptation of this selection
based on the detected clusters (e.g.\ a variable colour slope) could be made
in future releases.

The width of the colour slices are in part motivated by the effect of photometric
uncertainty: there will be a broadening in the sequence towards the faint end where
the photometric errors inflate. In a slice of fixed width, the contamination of
non-cluster members will increase, as will the rate of cluster members being randomly
scattered out of the slice. To this end, we enforce additional magnitude limits in
each band, set where the average 1$\sigma$ uncertainty in photometry becomes
comparable to the width of our colour slice. We define this to be the magnitude at
which 50\% of galaxies in the slice have errors equivalent to the width of the slice.
The limits in {\it griz} are 24.0, 23.5, 23.3 and 21.6\,mag. These cuts cull the input
catalogue to 3,346,380 galaxies.

The scan through colour-space is complete, in that we cover a parameter space
that should contain all red-sequences that could be detected in the optical
bands (i.e.\ before the 4000\AA\ break is red-shifted out of this window).
However, we do enforce a blue limit in $(g-r)$ which ensures we only select
colours redder than the $z=0$ red-sequence, which we derive by extrapolating
the equivalent colour at $r=20$\,mag from the sequences of Coma and Virgo
(Smith et al.\ 2009; Rines \& Geller\ 2008). This corresponds to $(g-r)>
0.47$\,mag, which fixes the bluest limit for a red-sequence in the catalogue.
Consecutive slices overlap, since in successive scans we increase the colour
normalisation of the selection slice by 0.04\,mag ($\sim$75\% overlap) and so
(by design) the same cluster may be detected more than once in different
selections. Since the red-sequence has some scatter (both natural and from
photometric errors), the contrast of the cluster against the background (after
colour selection) will rise to a peak and then vanish as the colour slice
moves red-wards of the ridge-line.

To improve the rejection of background and foreground sources, we filter the
catalogue using slices in two filters simultaneously, and we use two
combinations of filters to detect clusters over a wide redshift range: $\{
(g-r), (r-i) \}$ and $\{ (r-i), (i-z) \}$. To search for the highest redshift
clusters we can possibly detect using this technique, we also make a scan in
$(i-z)$ only. Multiple detections of the same cluster are identified if the
red-sequences (fit by a straight line) are within the width of the photometric
filter, or two clusters share the same BCG. To eliminate these multiple
detections, we pick the `best' cluster by selecting the cluster with the
largest `reduced flux': the sum of the flux of all members excluding the
brightest three galaxies. Murphy, Geach \& Bower\ (2010) describe in more
detail the exact procedure for merging multiple cluster detections into a
master catalogue.

\begin{figure*}
\centerline{\includegraphics[width=0.45\textwidth]{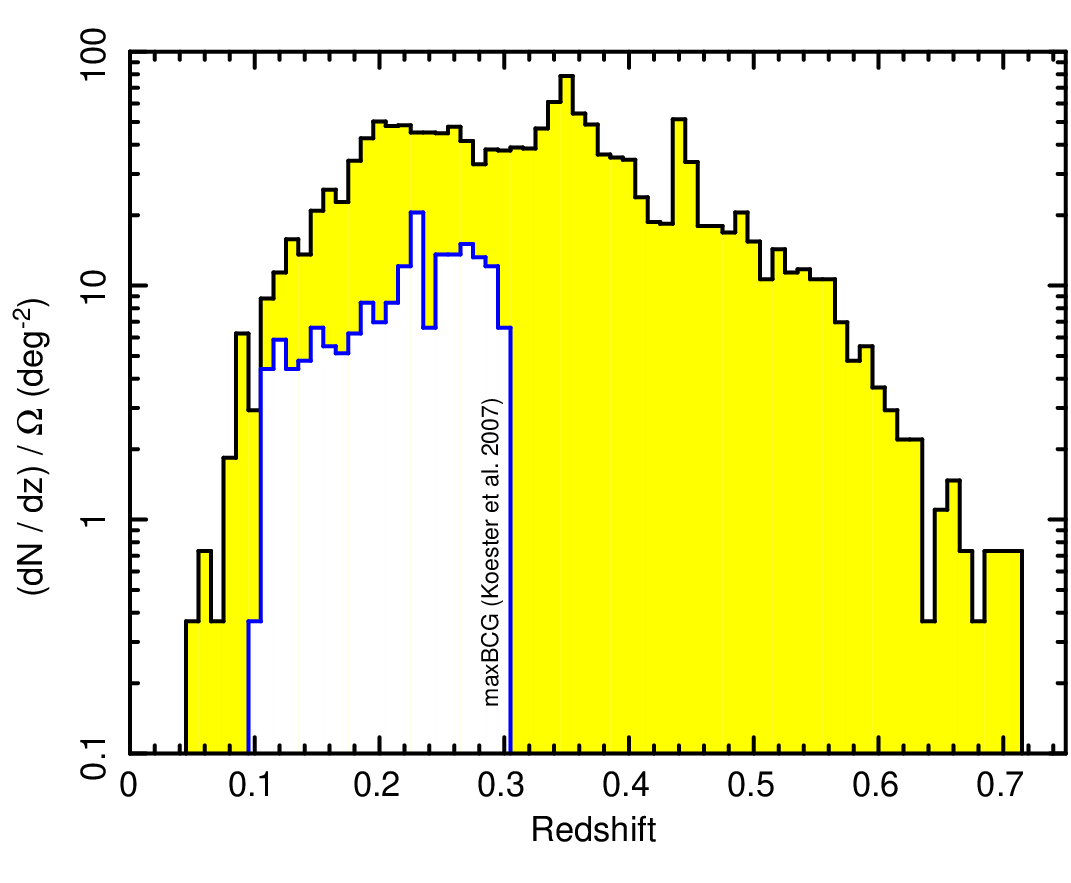}\includegraphics[width=0.45\textwidth]{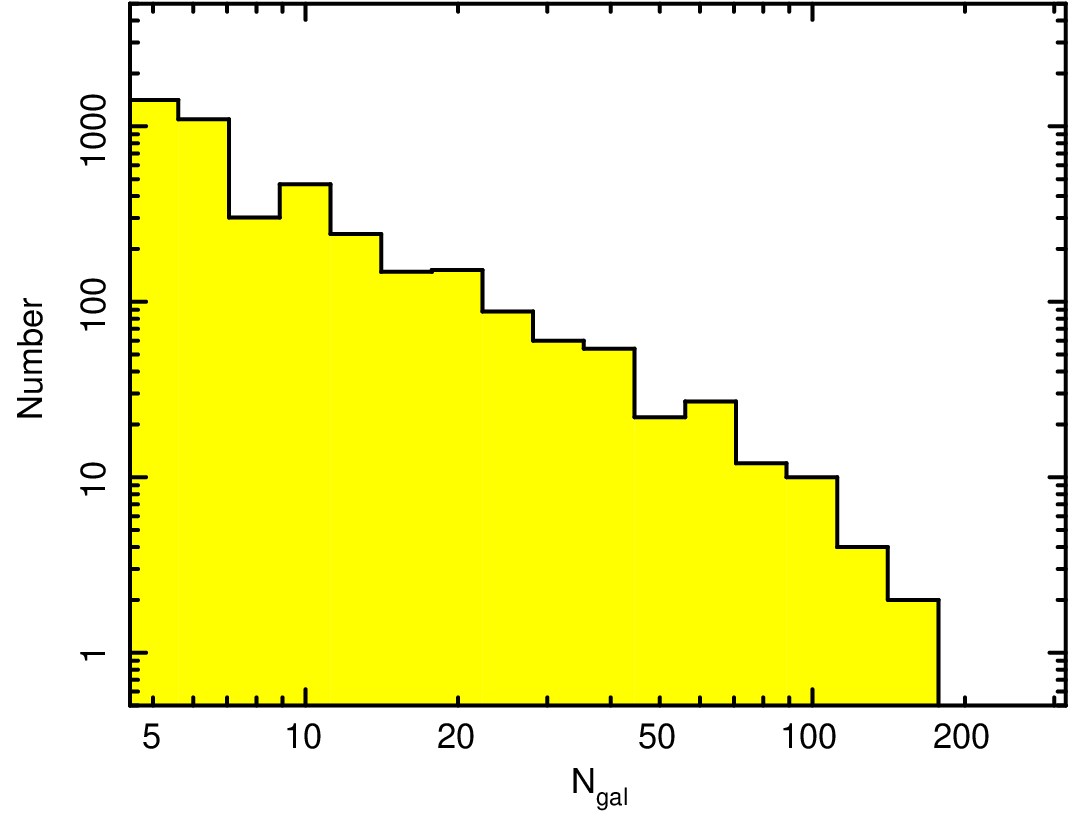}}
\centerline{\includegraphics[width=0.45\textwidth]{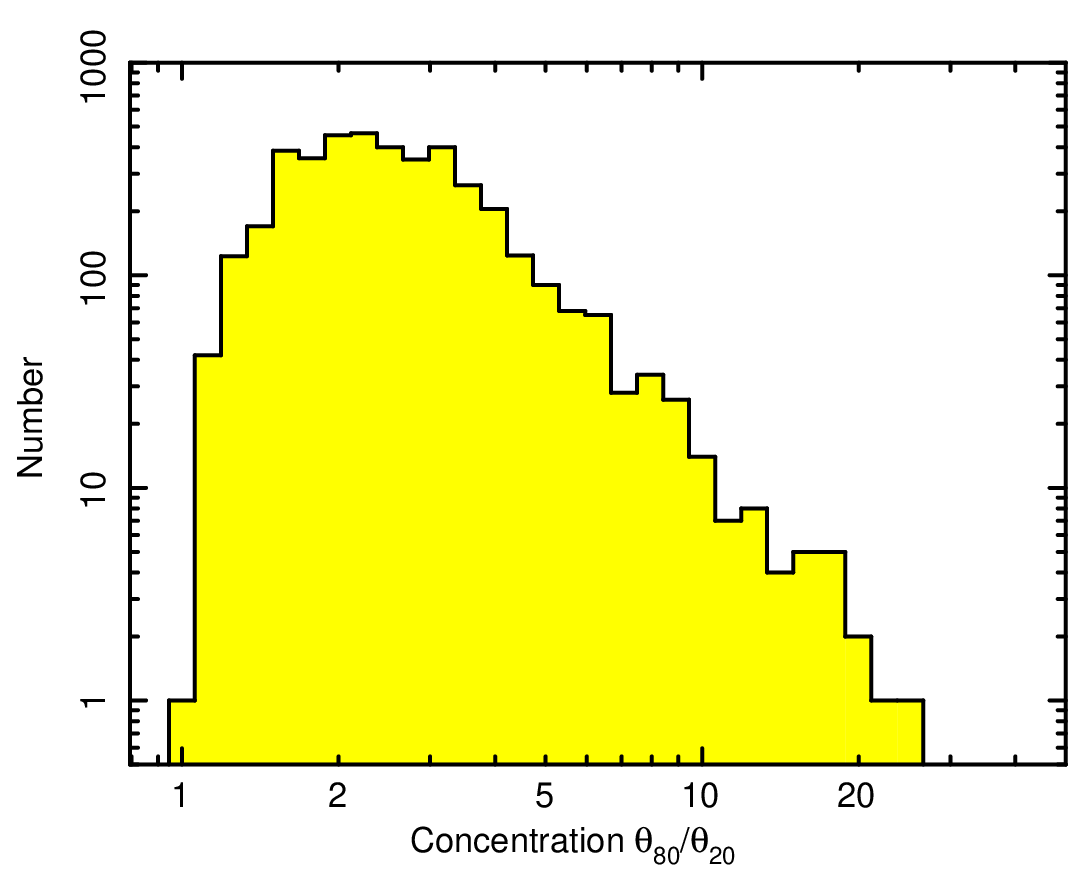}\includegraphics[width=0.45\textwidth]{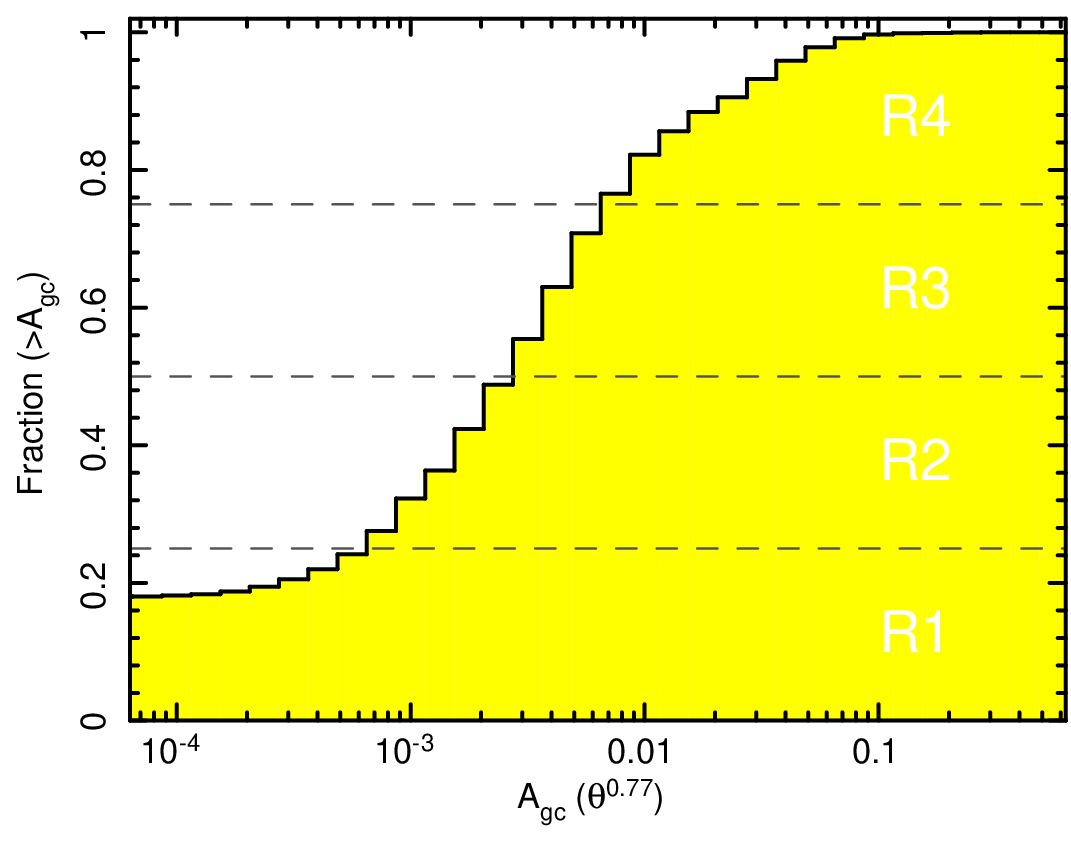}}
\caption{Vital statistics for the cluster catalogue. (top left)
  Redshift distribution for the Stripe 82 clusters, compared to the
  distribution of clusters identified using the {\it maxBCG} cluster
  detection algorithm (Koester et al.\ 2007a). Note that the {\it
    maxBCG} detector was not applied to the deeper multi-epoch
  photometry exploited in this work (see \S3.3). The median redshift
  of clusters in the present catalogue is $\left<z\right>=0.32$. (top
  right) The distribution of cluster membership, where the minimum
  criteria for a cluster was 5 members. (bottom left) Concentration
  distribution, $C$, defined as the ratio of the radii of apertures
  containing 80\% and 20\% of the members. The concentration could be
  a useful parameter for the selection of sub-sets of the
  clusters. (bottom right) Richness classification (see \S\S3.3.2),
  based on quartile ranges of the distribution of $A_{\rm gc}$ -- the
  estimated amplitude of the angular correlation function of galaxies
  in clusters. We define four simple classes of ascending richness:
  R1--R4.}
\end{figure*}

\subsection{Results}

After catalogue cleaning and duplicate rejection, we detect a total of
4098 unique clusters with $\geq$5 members. We have applied the same
algorithm to a mock catalogue generated from a semi-analytic
prescription for galaxy formation (Bower et al.\ 2006) within a
$\Lambda$CDM framework (the Millennium Simulation; Springel et al.\
2005) to assess the efficacy of the algorithm where we can evaluate the
completeness of the detection in terms of halo mass. This analysis
suggests that at $z<0.4$ we are $>$68\% complete for halos with
mass $\log(M_{\rm h}/h^{-1}M_\odot)\gtrsim13.6$. This rises to $>$90\%
completeness for the most massive halos with $\log(M_{\rm
  h}/h^{-1}M_\odot)\gtrsim14$. Full details of this analysis, along
with details of the mock catalogue are given in Murphy, Geach \& Bower
(2010).

It is important to note other cluster-finding efforts in the Stripe 82 region.
The most significant cluster catalogue comparable to this is arguably the catalogue of
clusters presented by Koester et al. (2007a) using the {\it maxBCG} algorithm (see
Koester et al.\ 2007b). These clusters were detected across the full SDSS footprint,
with 492 clusters in the Stripe 82 region. $\gtrsim$90\% of clusters in the {\it
maxBCG} catalogue (common to the Stripe 82 region) are detected in our catalogue.
Direct comparison of the relative efficacy of the two algorithms is unfair, since the
{\it maxBCG} catalogue was applied to the shallower SDSS photometry prior to the
multi-epoch Stripe 82 co-added data. As a result, we are able to detect more clusters
(including groups of fainter systems), out to higher redshifts than in Koester et al.\
(2007), as shown in Figure\ 2. Differences in the definition of a cluster also have an
important role, especially at the low mass, or poor, end of the sample. First,
restricting the current catalogue to $0.1\leq z\leq0.3$ to better match the range of
{\it maxBCG} we detect 1794 clusters and groups with $\geq$5 members. However, setting
the minimum membership to $\geq$10 members, we find 504 systems, much more comparable
to the {\it maxBCG} surface density.

Clearly then, the most important consideration that should be taken into account when
comparing cluster catalogues is the selection function at the low mass end. The most
massive clusters are likely to be detected easily even by very different techniques,
since these will generally be very prominent in projection. However, differences in
the completeness limit, minimum selection criteria and contamination rate can have a
significant impact. As a test, we consider a mock catalogue of clusters generated from
the Millennium Simulation, and populated with galaxies formed from the Bower et al.\
(2006) {\sc galform} recipe. The number of halos detected increases by a factor
$\sim$2 when between mass limits of 1--2$\times10^{14}M_\odot$ (the estimated
difference between the completeness limits of the {\it maxBCG} catalogue and ours).
These considerations should be taken into account when comparing optically-selected
cluster catalogues. Ideally, one would like to unify different catalogues that apply
different search techniques. Cross-calibrating such a catalogue with independent mass
estimates (such as weak lensing or X-ray luminosities), and tests on controlled
mock-catalogues could provide a powerful resource for investigating the nature of the
low-mass end of the cluster mass function.

In the following sub-sections we describe the main properties of our Stripe\ 82
cluster catalogue, including a description of our cluster redshift estimates, richness
evaluation and expected false detection rate. A sky-plot of clusters detected in
Stripe\ 82 is presented in Figure\ 1. Full details of the catalogue contents and
information on accessibility are given in Appendix\ 1.

\subsubsection{Redshift estimation}

Red-sequence galaxies lend themselves well to photometric redshift ($z_{\rm p}$)
estimation in the absence of spectroscopy: the prominent 4000\AA\ break serves as a
strong redshift discriminator in evolved galaxies. Approximately 32\% of clusters in
the catalogue have at least one member that has a spectroscopic redshift, while the
remainder of members have a photometric redshift. Since the Stripe 82 multi-epoch data
is deeper than the remainder of the SDSS DR7, we found some galaxies did not have
pre-computed photometric redshifts. To this end, we used the code {\it hyperz}
(Bolzonella et al.\ 2000) to estimate redshifts for any galaxy without an existing DR7
photometric redshift or spectroscopic redshift, exploiting the deeper {\it ugriz}
Stripe 82 photometry.

The dispersion in $\delta z/(1+z)$ for $z^{\rm hyperz}_{\rm p}$ in a
spectroscopically confirmed sample of 1549 galaxies in our cluster catalogue is 0.029,
compared to 0.017 for the same galaxies when the photometric redshift is calculated
with the DR7 algorithm (both figures calculated from the standard deviation in $\delta
z/(1+z)$ after rejecting galaxies with $>$3$\sigma$ clipping). We attribute the higher
precision in the DR7 photometric redshifts as due to the sophistication of the DR7
algorithm compared to our simple {\it hyperz} $\chi^2$ fits to a limited range of
spectral templates. Thus, at the median redshift of the cluster sample, we expect
photometric redshifts to be accurate to $\sim$5--9\%.

Since we have several different redshifts for a given cluster, we can combine this
information into a single redshift estimate for the cluster ensemble. We calculate a
weighted median redshift for the system; spectroscopic redshifts are given a weighting
of 4 (in effect, that galaxy is counted four times); photometric redshifts from the
DR7 catalogue have a weighting of 2, and the {\it hyperz} calculated redshifts are
given a weighting of unity. The lower weighting for the latter reflects the slightly
poorer performance of these photometric redshifts compared to those from DR7 described
above. A cone plot showing the redshift distribution of clusters in the Stripe is
shown in Figure\ 1, and a histogram of the redshift distribution is shown in Figure\
2. The median redshift of clusters in the survey is $\left<z\right>=0.32$, however the
depth of the multi-epoch SDSS data in this region allows us to detect clusters
comfortably out to $z\sim0.5$, with a handful of systems detected at $z\gtrsim 0.6$.

\subsubsection{Richness estimates}

Often it is convenient to classify clusters according to their `richness' --
i.e. an observable parameter that correlates with the mass of the structure.
In the absence of X-ray luminosities, velocity dispersions or accurate lens
models of the underlying matter profile, we must rely on cruder methods of
richness estimation that employ counting statistics to assess the significance
of the density enhancement in the cluster compared to the field.
Unfortunately, calibrating optical richness measurements to various mass
estimates across different surveys is notoriously difficult, and so in this
catalogue we have provided several (related) richness estimates based on
aperture counts of cluster members corrected for field contamination.
Moreover, the information provided in the cluster catalogue (Appendix\ 1)
should be sufficient for the reader to either re-calculate a specific richness
estimate, or re-calibrate our measured values to some other scale of their
choice.

All of our richness estimates are based on the net counts of galaxies within
an aperture of radius $\theta$ centred on the cluster centre (this is defined
as the geometric mean centre of all members, or the position of the brightest
cluster galaxy -- again, the reader can adopt either position accordingly):
\begin{equation} N_{\rm net} = N_T - N_B \end{equation} where $N_T$ is the
number of galaxies within the aperture, and $N_B$ is the number of background
galaxies selected in an annulus centred on the cluster, with equivalent area
to the $N_T$ selection aperture. Using an annulus instead of a scaled surface
density for the full catalogue, although resulting in poorer number
statistics, accounts for potential differences in photometric properties
(seeing, local extinction, etc.) across different regions of the stripe. We
have made no correction for the presence of bright star halos, or other
cosmetic effects that might affect the counts in apertures. We adopt two
values for $\theta$: \begin{enumerate} \item $\theta_{80}$ -- the radius of an
aperture containing 80\% of the members \item $\theta_{\rm 0.5Mpc}$ -- the
angular size of an aperture with projected physical size 0.5\,Mpc
\end{enumerate} Similarly, the counts can be defined as either (a) all
galaxies in the range $\left< m_3, m_3+3\right>$, where $m_3$ is the magnitude
of the third brightest cluster member, or (b) all galaxies in the photometric
slice (described in \S3.2) that the cluster was detected in. 

There are issues with both (a) and (b) that introduce uncertainty to the
richness calculation. We chose $m_3$ as a counting reference because it is
purely empirical and can easily be derived from the catalogue without any
additional assumptions about the cluster luminosity function. However, the
scatter in $m_3$ will inflate both for low-number and high-$z$ clusters due to
stochasticity, photometric uncertainties and projection effects. For example,
in a redshift slice $0.2<z<0.3$, the standard deviation of $m_3$ measured for
all clusters is strongly dependent on the number of galaxies assigned to the
cluster, $N_{\rm gal}$. For clusters with $5\leq N_{\rm gal}\leq10$ we find
$\sigma(m_3)=0.77$\,mag, dropping to $\sigma(m_3)=0.37$\,mag for richer
systems, $25\leq N_{\rm gal}\leq35$. Similarly, in case (b), counting galaxies
in a thin slice will result in uncertainty due to galaxies being scattered out
of and into the slice -- an issue that is also exacerbated for low-mass/faint
systems.

Despite their limitations, from these basic statistics, we can derive higher
order richness estimates. One commonly used measure is the $B_{\rm gc}$
statistic (Longair \& Seldner\ 1979; Yee \& Lopez-Cruz 2001) which has been
shown to scale well with other more direct measurements of cluster mass (Yee
\& Ellingson\ 2003). This statistic is designed to estimate the amplitude of
the spatial cross-correlation function for galaxies: \begin{equation}\xi(r) =
B_{\rm gc}r^{-\gamma}.\end{equation} To calculate $B_{\rm gc}$ requires a
de-projection of the amplitude of the {\it angular} correlation function into
3D space, and this is estimated by: \begin{equation} B_{\rm gc} =
\frac{\rho_gA_{\rm gc}}{\Phi(m_l,z)I_\gamma}d_{\theta}^{\gamma-3}
\end{equation} where $\rho_{\rm g}$ is the average surface density of galaxies
in the field brighter than a limit $m_l$, $d_\theta$ is the angular diameter
distance to the redshift of the cluster, $\gamma$ is the slope of the
power-law in the correlation function (equation\ 2), and $A_{\rm gc}$ is the
amplitude of the angular correlation function, estimated as: \begin{equation}
A_{\rm gc} = \frac{N_{\rm net}}{N_B}\frac{3-\gamma}{2}\theta^{\gamma-1}.
\end{equation} Finally, the $B_{\rm gc}$ statistic is scaled by the luminosity
function, integrated between the absolute magnitude of the second brightest
cluster member, down to the luminosity corresponding to $m_l$ at the redshift
of the cluster (note that $I_\gamma = 3.78$ -- an integration constant, and
$\gamma=1.77$). The limiting magnitude is set as $m_3+3$, where $m_3$ is the
third brightest member of the cluster. Both $B_{\rm gc}$ and $A_{\rm gc}$ for
both variants of angular scale and photometric selection described above are
provided in the cluster catalogue to be used at the reader's discretion, but
here we prefer the angular correlation function amplitude calculated for
$\theta_{80}$ and galaxies selected in the set of photometric filters the
cluster was detected in. Unlike $B_{\rm gc}$, here we require no scaling for
luminosity function. Finally, note that all of these statistics are ultimately
governed by Poisson noise in $N_T$ and $N_B$. Naturally, this leads to a
break-down of the practicality of these richness statistics in low-member
(group) systems, and so should only be taken as a guide.

In order to coarsely segregate the catalogue into richness bins, we define four
classifications of richness: R1--R4. These classifications are simply the quartile
ranges of the parameter $A_{\rm gc}$, calculated inside $\theta_{80}$ and for all
galaxies within the detection slice. The cumulative histogram and ranges are given in
Figure\ 2. Finally, taken with a richness estimate, the concentration of galaxies in
cluster can also be a useful parameter to describe the morphology of the system. We
define a simple dimensionless concentration parameter $C=(\theta_{\rm 80}/\theta_{\rm
20})$. Where $\theta_{\rm 20}$ and $\theta_{\rm 80}$ are the radii of a circular
aperture containing 20\% and 80\% of the members respectively. We find a median
concentration of $\left<C\right>\sim2.4$ (Figure\ 2).

\subsubsection{False detections and contamination}

A more comprehensive analysis of the completeness, purity, expected stellar mass
recovery, and other statistical measures of the performance of the algorithm making
use of mock catalogues are outlined in Murphy, Geach \& Bower\ (2010). However, here
it is instructive to outline the most important statistic pertinent to the current
catalogue: the expect false positive detection rate.

To evaluate the inclusion of `clusters' by erroneous random associations of
galaxies. To evaluate this, we take the original Stripe\ 82 catalogue of
galaxies and randomly shuffle the colours of galaxies, keeping the positions
the same. Keeping the positions of galaxies constant ensures we replicate the
natural random angular clustering of galaxies on the sky, but the
randomisation of the colours allows us to assess how often these random
associations could be linked by our assumption that group and cluster galaxies
will have very similar colours. We apply the algorithm in exactly the same
manner as the main detection, and find a rate of false detections of 0.06
`clusters' per square degree. The vast majority of these false detections are
made up of associations of 5 or 6 galaxies, near the lower cut-off for what we
consider a group. Thus, the rate of false detections caused by random
associations in the final catalogue is expected to be small, $<$1\%.

An important caveat is that there could be added contamination from other
incorrect identifications; mainly this will involve (i) associations of
`galaxies' that are actually the fragmented halos around bright stars; (ii)
heavily, uniformly reddened galaxies in regions of high Galactic extinction;
(iii) associations of `galaxies' from overly de-blended galaxies in the
catalogue. We have been careful to try to minimise the inclusion of such
systems in the original catalogue obtained from CAS, however it is possible
that some level of contamination in the final cluster catalogue could remain.
Therefore, while our survey was exhaustive over the entire Stripe 82 area, the
reader should exercise caution by flagging clusters that were detected in, for
example, the vicinity of bright stars.

Another important source of contamination and completeness is the issue of the
erroneous merging of structures along the line of sight, close in redshift
(and therefore colour) space that are not physically connected. The separation
of these systems in our algorithm depends on the relative difference between
the red-sequences compared to the width of the colour `slices' the two systems
were detected in. To investigate how this might affect our catalogue, we have
artificially created systems that are aligned along the line of sight and have
identical red sequences. The colour of one of the sequences is reddened until
the detector resolves the clusters into a pair; this indicates the minimum
separation in redshift space (assuming a simple stellar population model for
the colour difference of the red galaxies) at which the clusters can be
resolved. In all cases, the projected clusters can be resolved when the
separation between the red-sequences is approximately half of the width of the
colour slice (of order 0.1\,mag, see \S3.2). Structures along the line of
sight that have colour sequences separated by less than this are grouped into
a single system and will only be disentangled by follow-up spectroscopy which
can determine the relative velocity offsets of potential merged systems.

On a related note, the discrimination of structures at the same redshift, but
separated by some projected distance is also important. The choice of critical
threshold for Voronoi cell area (\S2.1) can result in `valleys' in the surface
density of galaxies that could potentially fragment clusters with large
amounts of sub-structure (e.g.\ two dense cores connected by some lower
density filamentary structure). Based on experiments with mock catalogues, we
have attempted to optimise the algorithm such that the level of fragmentation
does not over-split clusters, whilst maintaining acceptable levels of cluster
`purity', completeness, etc. Full details of these experiments can be found in
a sister paper, Murphy, Geach \& Bower\ (2010).

\subsection{Quasar sight-line catalogue}

A powerful observational technique is the exploitation of continuum-bright
background sources to search for spectroscopic evidence of absorption of
continuum light by intervening intergalactic/intracluster material. For
example, Lopez et al.\ (2008) present a study of 442 cluster-quasar pairs
(sight-lines) within the Red Sequence Cluster Survey (RCS; Gladders \& Yee
2000, 2005) where the Mg\,{\sc ii} $\lambda\lambda$2796,2803 doublet could be
detected at $0.3<z<0.9$. The study of the distribution of the equivalent width
of absorption-line systems within clusters could provide a means of studying
environmental effects such as gas-stripping in dense environment. The
potential identification of high-ionisation absorption-line systems in the
X-ray and UV could also provide a window onto the
warm-hot phase of the intergalactic medium.

The large number of quasars already catalogued in the SDSS provides us with
the opportunity to identify further potential targets for future sight-line
studies, and here we supply a simple supplementary catalogue of 18,295 QSO-cluster
pairs that might be useful for this purpose. We have taken the catalogue of
Veron et al.\ (2010) and identified all QSOs with $0.25<z<2$ and
$V\leq21$\,mag within a projected radius of foreground clusters that
corresponds to 3 proper Mpc at the cluster redshift. A sub-set of the catalogue is given in Appendix\ 1 as a guide for content, and the full catalogue is available online at www.physics.mcgill.ca/$\sim$jimgeach/stripe82.

\section{Summary}  

We have presented a catalogue of 4098 photometrically detected galaxy
clusters in the Sloan Digital Sky Survey `Stripe 82' equatorial
multi-epoch co-add, a $\sim$270 square degree strip with photometry
$\sim$2\,mags deeper than the general SDSS imaging survey. The
clusters have a median redshift of $\left<z\right>=0.32$, and we can
comfortably detect systems out to $z\sim0.5$, although photometry in
redder bands is required for the efficient detection of
higher-redshift systems. In addition to the cluster catalogue, we
provide a supplementary catalogue of 18,295 $V\leq21$\,mag background
QSO sight-lines, all within a projected radius of 3 proper Mpc of
foreground clusters in this catalogue. These QSOs are simply
cross-matches between clusters and QSOs in the catalogue of Veron et
al.\ (2010). The sight-line catalogue will be a useful resource for
future follow-up spectroscopic studies whose goals are the study of
(for example) absorption line systems in cluster environments.

This catalogue is publicly available, and will be maintained from
http://www.physics.mccgill.ca/$\sim$jimgeach/stripe82. Full details on access and
content of the catalogue given in the Appendix of this paper. Readers are encouraged
to contact the authors for any further information or assistance with the catalogues.
We expect to improve the catalogue in future releases, with follow-up imaging and
spectroscopy, and refinements of the detection algorithm. A forthcoming publication
will present a more extensive catalogue of galaxy clusters detected using the same
technique using SDSS DR7 data across the full SDSS footprint.

\section*{acknowledgements} We thank the anonymous referee for insightful
comments that have improved this paper. The authors thank the National
Science and Engineering Research Council of Canada and the U.K. Science and
Technology Facilities Council for financial support. We are also
indebted to Neil Crighton and Britt
Lundgren for assistance with the QSO sight-line catalogue and Alastair Edge and
Ian Smail for helpful discussions.

\onecolumn

\section*{Appendix: distribution of the catalogue}
The cluster catalogue is available at
http://www.physics.mcgill.ca/$\sim$jimgeach/stripe82.

\smallskip

\noindent We have chosen to distribute the full catalogue in
Hierarchical Data Format (HDF, version 5). This format provides a
natural way for us to release both the `top level' cluster properties
(co-ordinate, redshift, etc.)  {\it and} a wide range of other
information, including various richness classifications, and
information for each of the constituent galaxies. However, for simple
access, we also provide a simple FITS table with just the basic
`top-level' data in.  In Table\ A1 we list and describe the HDF data
structure of the cluster catalogue. Both HDF5 and FITS format
catalogues products are available for download at the website listed
above.

For the QSO sight-line supplementary catalogue, we provide a simple
FITS standard file with information on the QSO itself (taken from
Veron et al.\ 2010), and a link to the relevant cluster in the main
catalogue; however for convenience, we also provide some basic data on
the matching cluster in this catalogue. The contents of the sight-line
catalogue is provided in Table\ A2, and the file is available also
from the website.

\clearpage

\begin{table*}
  \caption{Contents and hierarchy description of the Stripe 82 cluster
    catalogue. The top level information is also available as a
    stand-alone FITS table.}
\begin{tabular*}{0.99\textwidth}{@{\extracolsep{\fill}}ll}
  \hline
  Hierarchy & Description\cr
  
  \cr
  \hline
  \multicolumn{2}{c}{Top level}\cr
  \hline
  {\tt /ClusterNNNNN/ID} & Cluster identification number \cr
  {\tt /ClusterNNNNN/ra} & Cluster right ascension (degrees, J2000) \cr
  {\tt /ClusterNNNNN/dec} & Cluster declination (degrees, J2000)\cr
  {\tt /ClusterNNNNN/ra\_bcg} & BCG right ascension (degrees, J2000) \cr
  {\tt /ClusterNNNNN/dec\_bcg} & BCG declination (degrees, J2000)\cr
  {\tt /ClusterNNNNN/ngal} & Number of galaxies assigned to cluster \cr
  {\tt /ClusterNNNNN/redshift} & Cluster redshift \cr
  {\tt /ClusterNNNNN/redshift\_code} & Code
  describing composite average of galaxy redshifts \cr
  {\tt /ClusterNNNNN/theta80} & Angular radius containing
  80\% of the members ($\theta_{80}$)\cr
  {\tt /ClusterNNNNN/concentration} & $\theta_{80}/\theta_{20}$
  concentration measurement \cr
{\tt /ClusterNNNNN/Agc} & Richness estimator $A_{\rm gc}$\cr
  \hline
  \multicolumn{2}{c}{Richness measurements}\cr
  \hline
  {\tt /ClusterNNNNN/Richness/Class} & Richness class R1--R4\cr
  {\tt /ClusterNNNNN/Richness/R500/RedSequence/...} & Measured within
  0.5\,Mpc using red-sequence selection\cr
  {\tt /ClusterNNNNN/Richness/R500/MagLim/...} & Measured within
  0.5\,Mpc using magnitude limited selection\cr
  {\tt /ClusterNNNNN/Richness/Theta80/RedSequence/...} &  Measured within
  $\theta_{80}$ using red-sequence selection \cr
  {\tt /ClusterNNNNN/Richness/Theta80/MagLim/...} & Measured within
  $\theta_{80}$  using magnitude limited selection \cr

  {\tt .../Nb} & Number of background
  galaxies within aperture\cr
  {\tt .../Nnet} & Net number of galaxies within aperture\cr
  {\tt .../Agc} & Richness estimator $A_{\rm gc}$\cr
  {\tt .../Bgc} & Richness estimator $B_{\rm
    gc}$ \cr
  \hline
  \multicolumn{2}{c}{Galaxy member information}\cr
  \hline
  {\tt /ClusterNNNNN/Galaxies/GalaxyNNN/objID} & Member galaxy SDSS
  Stripe 82 {\tt PhotObjID} \cr
  {\tt /ClusterNNNNN/Galaxies/GalaxyNNN/DR7\_objID} & Member galaxy
  SDSS DR7 {\tt PhotObjID} \cr
  {\tt /ClusterNNNNN/Galaxies/GalaxyNNN/ra} & Member galaxy $N$ right ascension
  (degrees, J2000)\cr
  {\tt /ClusterNNNNN/Galaxies/GalaxyNNN/dec} & Member galaxy $N$  declination  (degrees, J2000)\cr
  {\tt /ClusterNNNNN/Galaxies/GalaxyNNN/specz} & Member galaxy $N$
  spectroscopic redshift \cr
  {\tt /ClusterNNNNN/Galaxies/GalaxyNNN/specz\_source} & Member galaxy $N$
  spectroscopic redshift source\cr
  {\tt /ClusterNNNNN/Galaxies/GalaxyNNN/photoz} & Member galaxy $N$ photometric redshift \cr
  {\tt /ClusterNNNNN/Galaxies/GalaxyNNN/photoz\_source} & Member galaxy $N$
  photometric redshift source\cr
  {\tt /ClusterNNNNN/Galaxies/GalaxyNNN/u} & Member galaxy $N$  $u$-band model mag \cr
  {\tt /ClusterNNNNN/Galaxies/GalaxyNNN/g} & Member galaxy $N$  $g$-band model mag \cr
  {\tt /ClusterNNNNN/Galaxies/GalaxyNNN/r} & Member galaxy $N$  $r$-band model mag \cr
  {\tt /ClusterNNNNN/Galaxies/GalaxyNNN/i} & Member galaxy $N$  $i$-band model mag \cr
  {\tt /ClusterNNNNN/Galaxies/GalaxyNNN/z} & Member galaxy $N$
  $z$-band model mag \cr
  \hline
\end{tabular*}
\end{table*}

\begin{table*}
  \caption{Contents and brief sub-set of the supplementary QSO
    sight-line catalogue. QSO ID and information is taken from the
    catalogue of Veron et al.\ (2010). The cluster ID corresponds to
    the ID listed in Table A1, although the corresponding cluster coordinates and
    redshift are also listed in this table. Note that one cluster may
    have multiple sight-lines. The impact parameter of the QSO to the
    cluster co-ordinate is given in terms of angle ($\theta$) and
    distance at the cluster redshift ($R$).}
\begin{tabular*}{0.99\textwidth}{@{\extracolsep{\fill}}llccccccccc}
  \hline
  QSO ID & Cluster ID &  QSO $z$ & QSO $\alpha$ & QSO $\delta$  & QSO
  $V$& $\theta$ & $R$ & Cluster $z$ &
  Cluster $\alpha$  & Cluster $\delta$ \cr
  & & & (deg) & (deg) & (mag) & (deg) & (Mpc) & & (deg) & (deg)\cr
  \hline
  119625 & 3390832343296285895 & 1.534 & 309.36000 & $-$0.346389 & 20.37 & 0.117 & 2.22 & 0.27 & 309.470087 & $-$0.306428\\
  165873 & 3390832343296285895 & 0.634 & 309.47875 & $-$0.472500 & 20.26 & 0.166 & 3.16 & 0.27 & 309.470087 & $-$0.306428\\
  119731 & 2567712495232746752 & 0.397 & 310.47292 & 0.485833 & 18.72 & 0.246 & 3.04 & 0.17 & 310.715893 & 0.521082\\
  119752 & 2567712495232746752 & 1.378 & 310.63167 & 0.744444 & 19.51 & 0.239 & 2.96 & 0.17 & 310.715893 & 0.521082\\
  119775 & 2567712495232746752 & 1.001 & 310.78667 & 0.781667 & 19.71 & 0.270 & 3.35 & 0.17 & 310.715893 & 0.521082\\
  119785 & 2567712495232746752 & 1.915 & 310.86667 & 0.635000 & 20.09 & 0.189 & 2.34 & 0.17 & 310.715893 & 0.521082\\
  165896 & 2567712495232746752 & 0.317 & 310.91667 & 0.481389 & 18.92 & 0.205 & 2.54 & 0.17 & 310.715893 & 0.521082\\
  \hline
\end{tabular*}
\end{table*}

\end{document}